\documentclass[11pt,a4paper]{article}
\usepackage{jheppub}
\usepackage{graphicx}
\usepackage{epsfig}
\usepackage{amsmath, amssymb}
\usepackage{dcolumn}
\usepackage{bm}
\usepackage{amsfonts}
\usepackage[toc,page]{appendix} 
\usepackage{color}

\usepackage{subfigure}
\usepackage{amssymb}
\usepackage{epstopdf}

\newcommand{\bea}[1]{\begin{eqnarray} \mbox{$\label{#1}$}}
\newcommand{\eea}{\end{eqnarray}}
\newcommand{\be}[1]{\begin{equation} \mbox{$\label{#1}$}}
\newcommand{\ee}{\vspace{0.1cm}\end{equation}}

\newcommand{\bi}{\begin{itemize}}
\newcommand{\ei}{\end{itemize}}

\newcommand{\TT}{{\rm TT}}

\newcommand{\bk}{{\mathbf k}}

\newcommand{\bx}{{\mathbf x}}

\newcommand{\de}{\delta}
\newcommand{\HH}{{\cal H}}

\begin{document}

\title{Can Self-Ordering Scalar Fields explain the BICEP2 B-mode signal?}

\author[a]{Ruth Durrer}
\emailAdd{ruth.durrer@unige.ch}
\author[a]{Daniel G. Figueroa}
\affiliation[a]{D\'epartement de Physique Th\'eorique and Center for Astroparticle Physics, Universit\'e de Gen\`eve, 24 quai Ernest Ansermet, CH--1211 Gen\`eve 4, Switzerland}
\emailAdd{daniel.figueroa@unige.ch}
\author[a,b]{Martin Kunz}
\emailAdd{martin.kunz@unige.ch}
\affiliation[b]{African Institute for Mathematical Sciences, 6 Melrose Road, Muizenberg,
7945, South Africa}

\date{\today}


\abstract{We show that self-ordering scalar fields (SOSF), i.e.~non-topological cosmic defects arising after a global phase transition, {\em cannot} explain the B-mode signal recently announced by BICEP2. We compute the full $C_\ell^{B}$ angular power spectrum of B-modes due to the vector and tensor perturbations of SOSF, modeled in the large-$N$ limit of a spontaneous broken global $O(N)$  symmetry. We conclude that the low-$\ell$ multipoles detected by BICEP2 cannot be due mainly to SOSF, since they have the wrong spectrum at low multipoles. As a byproduct we derive the first cosmological constraints on this model, showing that the BICEP2 B-mode polarization data admits at most a 2-3\% contribution from SOSF in the temperature anisotropies, similar to (but somewhat tighter than) the recently studied case of cosmic strings.
}

\maketitle

\section{Introduction}
\label{sec:intro}

The anisotropies of the cosmic microwave background (CMB) have been measured very precisely over the years by a range of different experiments~\cite{Larson:2010gs,Komatsu:2010fb,Dunkley:2010ge,Reichardt:2011yv,Ade:2013ktc,Planck:2013kta}, most recently by the {\it Planck} Satellite~\cite{Planck:2013kta}, whose temperature data demonstrate an impressive agreement with the standard flat $\Lambda$CDM model over all angular scales relevant at decoupling. Quite surprinsingly, the BICEP2 collaboration~\cite{BICEP2} has recently announced the first detection of a B-mode polarization signal at large angular scales. If confirmed, this detection opens a new observational window to models of the early Universe. 

The leading candidate to explain the BICEP2 low-$\ell$ B-mode signal are tensor perturbations generated during inflation due to quantum fluctuations of the metric. Inflationary tensor modes with a tensor to scalar ratio $r \simeq 0.2$ fit well the observed B-mode angular spectrum at scales $\ell \simeq 50-140$. The higher $\ell$ signal, dominated by the B-modes generated by the lensing of E-modes, shows some excess compared to the expected amplitude. However, the BICEP2 collaboration has warned us~\cite{BICEP2} that data at $\ell \gtrsim 150$ should be considered as preliminary.

A  mechanism alternative to inflation which generates primordial B-modes are cosmic defects, which may have have formed after a symmetry breaking phase transition in the early Universe~\cite{Kibble:1976sj,Kibble:1980mv}. Defects can be {\em local} or {\em global}, depending on whether  they are generated after a phase transition which breaks a gauge or a global symmetry. In the local case only cosmic strings are cosmologically viable defects leaving an imprint in the CMB. In the global case all defects are viable independent of their dimension, except for domain walls which are not allowed. For reviews on cosmic defects see~\cite{VilenkinAndShellard,HindmarshAndKibble,Durrer:2001cg,Hindmarsh:2011qj}. Cosmic defects lead to a variety of phenomenological effects, in particular they produce CMB temperature and polarization anisotropies~\cite{Durrer:1994zza,Durrer2,Durrer3,Turok1,Turok2}, which in general are expected to be non-Gaussian~\cite{Mark1,Mark2,Shellard,Dani}. 

Backgrounds of gravitational waves (GW) are also expected from several different processes related to cosmic defects: the creation~\cite{GarciaBellido:2007dg,GarciaBellido:2007af,Dufaux:2007pt,DufauxFigueroaBellido,Buchmuller:2013lra}, evolution~\cite{Krauss:1991qu,JonesSmith:2007ne,Fenu:2009qf, Giblin:2011yh,Figueroa:2012kw} and the decay~\cite{Vilenkin:1981bx,Vachaspati:1984gt, Olmez2010,Blanco-Pillado:2013qja} (the latter only applies to cosmic strings).
Though in general GW are expected to generate B-mode polarization in the CMB, not every background of GW can lead to  a signal at the relevant CMB scales. Essentially, creating low-$\ell$ B-mode polarization requires tensor modes with a significant amplitude at super-horizon scales at the time of decoupling. The GW backgrounds from the formation and decay of cosmic defects, are not expected to contribute a significant B-signal in the low multipole $\ell \simeq \mathcal{O}(10)-\mathcal{O}(100)$ range, simply because their spectra have power mainly at smaller (sub-horizon) scales, i.e.~larger $\ell$. Tensor perturbations created during the evolution of a scaling defect network will contribute to create B-mode polarization patterns mostly at smaller angular scales, however they will also have some power at large angular scales.

At late time, the GW spectrum produced during the evolution of a defect network is exactly scale-invariant~\cite{Figueroa:2012kw}, so one may wonder whether it is possible to distinguish it from a flat (i.e.~zero tilt) inflationary tensor spectrum. Would it produce a similar B-mode pattern in the polarization of the CMB as the one expected from inflationary flat tensor modes? Fortunately, a possible confusion between the two GW backgrounds only concerns the direct detection of GW by interferometers: the scale-invariance of tensor perturbations from the evolution of a defect network is only obtained once the modes have entered the horizon during the radiation dominated era.  At super horizon scales, however, in a   radiation or matter dominated Universe, the GW spectrum from defects is white noise, simply dictated by the spectrum of the source [see below Eq.~(\ref{eq:SuperHtentors})]. A low-$\ell$ B-signal in the CMB due to tensors, on the other hand, concerns modes which either just crossed the horizon or are still super horizon at the time of decoupling. A polarization B-signal in the CMB at large angular scales from tensor perturbations of cosmic defects is in fact expected to be quite different from the corresponding signal of inflationary tensor perturbations\footnote{Note, however, that the temperature angular spectrum from cosmic defects (and scaling seeds in general) on large angular scales which are super horizon at decoupling, is scale invariant. This can be understood by purely dimensional arguments simply because it is dominated by the integrated Sachs Wolfe effect~\cite{Durrer:1996va}.}.
 
Moreover, cosmic defects not only actively create  tensor perturbations during their evolution, but also vector and scalar perturbations. Vector perturbations also source B-modes and therefore, in order to properly asses the possible contribution from cosmic defects into the B-mode polarization of the CMB, it is necessary to consider the contribution from both tensor and vector perturbations.

After the BICEP2 announcement~\cite{BICEP2} two independent groups~\cite{Lizarraga:2014eaa,Moss:2014cra} have analyzed the possible contribution from cosmic strings to the B-mode polarization. Although different modeling of the string networks were used, both groups conclude that standard cosmic strings can contribute at most a small fraction of few $\%$ to the BICEP2 B-mode signal. Given the current upper bounds set by {\it Planck}~\cite{Ade:2013xla} on the fractional contribution $f_{10}$ from cosmic defects into the temperature angular power spectrum at multipole $\ell = 10$, 
it is easy to guess that no significant contribution from cosmic defects to the B-mode polarization signal detected by BICEP2 can be expected. This has been shown explicitely by ~\cite{Lizarraga:2014eaa,Moss:2014cra} for the case of local strings (\cite{Lizarraga:2014eaa} also considered semi-local strings and textures), and indeed it was anticipated by~\cite{Urrestilla:2008jv} years ago. 

In this paper we carry out a similar analysis to~\cite{Lizarraga:2014eaa,Moss:2014cra} but for a different type of cosmic defects, often referred to as {\it self-ordering~scalar~field} (SOSF), corresponding to non-topological configurations of global fields. The analysis of~\cite{Moss:2014cra}  indicates that a peculiar type of heavy strings (very different from the local ones) could in principle explain the BICEP2 signal without the need of inflation. The spectra of such a network of rare strings could resemble that of global defects, suggesting that SOSF might improve the fit to the BICEP2 data. A brief analysis which appeared immediately after the BICEP2 announcement~\cite{Dent:2014rga}, even concluded that SOSF could fully explain the unexpectedly large BICEP2 B-mode signal. On the other hand, \cite{Lizarraga:2014eaa} argued that the similarity of the B-mode polarization signal from cosmic strings, semilocal strings and texture means that none of these models can fit the data (as shown in Fig. 3 of~\cite{Lizarraga:2014eaa} for the texture case). Here we review these claims, using the most extreme SOSF model (the large-$N$ limit), clarifying certain aspects about SOSF, and incorporating the most precise calculations available of the tensor and vector metric perturbations in this scenario. We shall come to the same conclusion as \cite{Lizarraga:2014eaa}, namely that SOSF can contribute at best a few percent to the BICEP2 signal.

In section~\ref{sec:SOSF} we briefly review the basics of SOSF and their modelling in the large-$N$ limit of a global $O(N)$-model. In section~\ref{sec:BBspectra} we clarify certain aspects of SOSF, and compute the full $C_\ell^{B}$ angular power spectrum of B-modes due to the vector and tensor perturbations from SOSF. We then compare with the {\it Planck} temperature and BICEP2 polarization data. In section~\ref{sec:conclusions} we discuss the implications of our results and conclude.

\section{Self Ordering Scalar Fields. The large-N limit}\label{sec:SOSF}

In the early Universe, a phase transition due to the spontaneous breaking of a global $O(N)$ symmetry into $O(N-1)$, leads to the a network of cosmic defects. These are strings, monopoles or textures if $N = 2, 3$ or $4$, respectively (the case of domain walls $N = 1$ is not cosmologically viable). The case of "non-topological textures", $N > 4$, corresponds to non-topological field configurations, described by an N-component scalar field $\Phi^\dagger = {1\over\sqrt{2}}(\phi_1,\phi_2,...,\phi_N)$, which  remains in the vacuum manifold,  $\Phi^\dagger\Phi = {1\over2}\sum_a\phi_a^2({\bf x},t) = {1\over2}v^2$, after the phase transition, here $v$ denotes {\it vacuum~expectation~value}.  On co-moving distances larger than the the co-moving Hubble radius, $|\bx-\bx'|>\HH^{-1}$, the direction of $\Phi(\bx,t)$ and $\Phi(\bx',t)$ within the vacuum manifold are uncorrelated due to causality. This leads to a gradient energy density associated to the $N-1$ Goldstone modes of the field, $\rho \sim (\nabla\Phi)^2$. As the universe evolves the field 'self-orders' to minimize its energy, adopting at every time a configuration correlated inside causal regions of size of the order of the Hubble radius $\sim \HH^{-1}$, while the relative orientation at larger distances remains random. This self-ordering process, which continues indefinitely as the Hubble radius grows in time, is a manifestation of the so called {\it scaling} property of cosmic defects: the energy density of a network of defects scales with the expansion of the universe such that the correlation length is always of the order of the causality scale (the Hubble scale), and the fractional contribution to the total energy budget of the universe remains marginal.

If $N \gg 1$, i.e.~in the large-$N$ limit of a global $O(N)$ symmetric scalar field, the equations of motion of the field components after the completion of the symmetry breaking, can be linearized and solved exactly up to corrections of order $1/N$~\cite{Turok:1991qq}, yielding
\begin{eqnarray}\label{eq:LargeNapprox}
\phi_a(k,t) = \sqrt{A}\left(\frac{t}{t_*}\right)^{{3\over2}-\nu}
\frac{J_\nu(y)}{(y_*)^\nu} \,\phi_a(k,t_*)\,,
\end{eqnarray}
with $y=kt$, $y_*=kt_*$ ($t$ the conformal time), $\nu = 2$ and $A = 5\pi/4 \simeq 3.930$ in the radiation era, $\nu = 3$ and $A = 945\pi/32 \simeq 92.775$ in the matter eta, and $\phi_a(k,t_*)$ the $a$-th component of the field at the initial time $t_*$ (end of the phase transition). In the large-$N$ limit, $\phi_a$ is initially distributed with a white noise spectrum on large scales and vanishing power on small scales $\langle\phi_a(\bk,t_*)\phi_b(\bk',t_*)\rangle = (2\pi)^3\de_{ab}\frac{v^2}{N}\Theta(1-kt_*)\de(\bk+\bk')$, with $\Theta(x)$ the Heaviside step function. 
The field is aligned on scales smaller than the comoving horizon $\HH_*^{-1} \sim  t_*$, and has arbitrary orientation on larger scales. This allows for an analytical understanding of the evolution of the resulting non-topological field configurations. In addition, the calculation of the field energy-momentum tensor $T_{\mu\nu}(\lbrace \phi_a \rbrace)$ and its unequal time correlators (UTC), $\langle T_{\mu\nu}(\bk,t)T_{\alpha\beta}(\bk',t') \rangle \equiv (2\pi)^3U_{\mu\nu\alpha\beta}(kt,kt')\delta(\bk'+\bk)$, only requires convolution integrals and no expensive numerical lattice simulations, see~\cite{Durrer:2001cg,Fenu:2009qf,Fenu:2013tea} for more details.  

From now on we will refer to this approximation simply as the large-$N$ scenario, or simply as the large-$N$. The large-$N$ UTC's, computed with a highly  accurate integrator over a large range of scales, have been recently used to compute all CMB temperature and polarization anisotropies~\cite{Fenu:2013tea}. In particular, isolating the vector and tensor degrees of freedom of the large-$N$ energy-momentum tensor $T_{\mu\nu}$, the corresponding vector and tensor UTC's, say $U_v(y,y')$, $U_t(y,y')$, were obtained for scales spanning 6 orders of magnitude around the horizon scale, $10^{-3} \le  y,y' \le 10^3$. After diagonalizing each UTC, $U(y,y') = \sum_a \lambda_a v_a(y)v_a(y')$, the eigenvectors were used as an active source of vector/tensor perturbations in a Boltzmann integrator, from which the B-mode polarization angular power spectra $C_{\ell}^B =C_{\ell}^{B(v)} + C_{\ell}^{B(t)}$ are obtained. The details of  this are rather technical, and we refer the interested reader to Ref.~\cite{Fenu:2013tea}.

\section{B-mode polarization signal from the large-N limit of SOSF}
\label{sec:BBspectra}

\begin{figure}[t]
\begin{center}
\includegraphics[width=12cm]{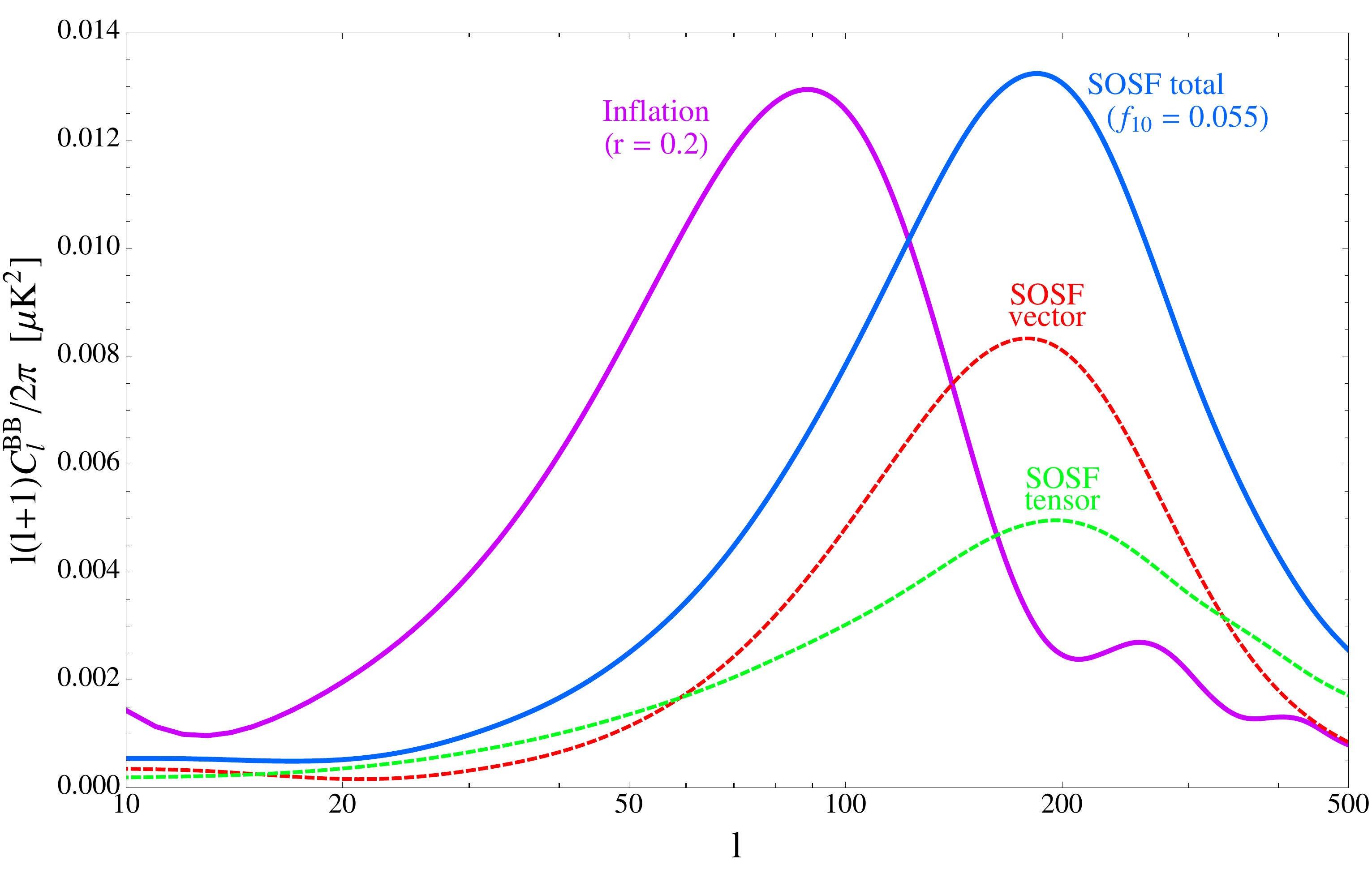}
\end{center}
\caption{B-mode angular power from SOSF for $f_{10} = 0.055$ without lensing contribution (blue continuous line), which is the sum of the vector (red dashed line) and tensor (green dashed line) contributions. The B-mode angular power from inflation for $r =0.2$ also without lensing is shown (purple continuous line) for comparison. }
\label{fig:1}
\end{figure}

\begin{figure}[t]
\begin{center}
\includegraphics[width=12cm]{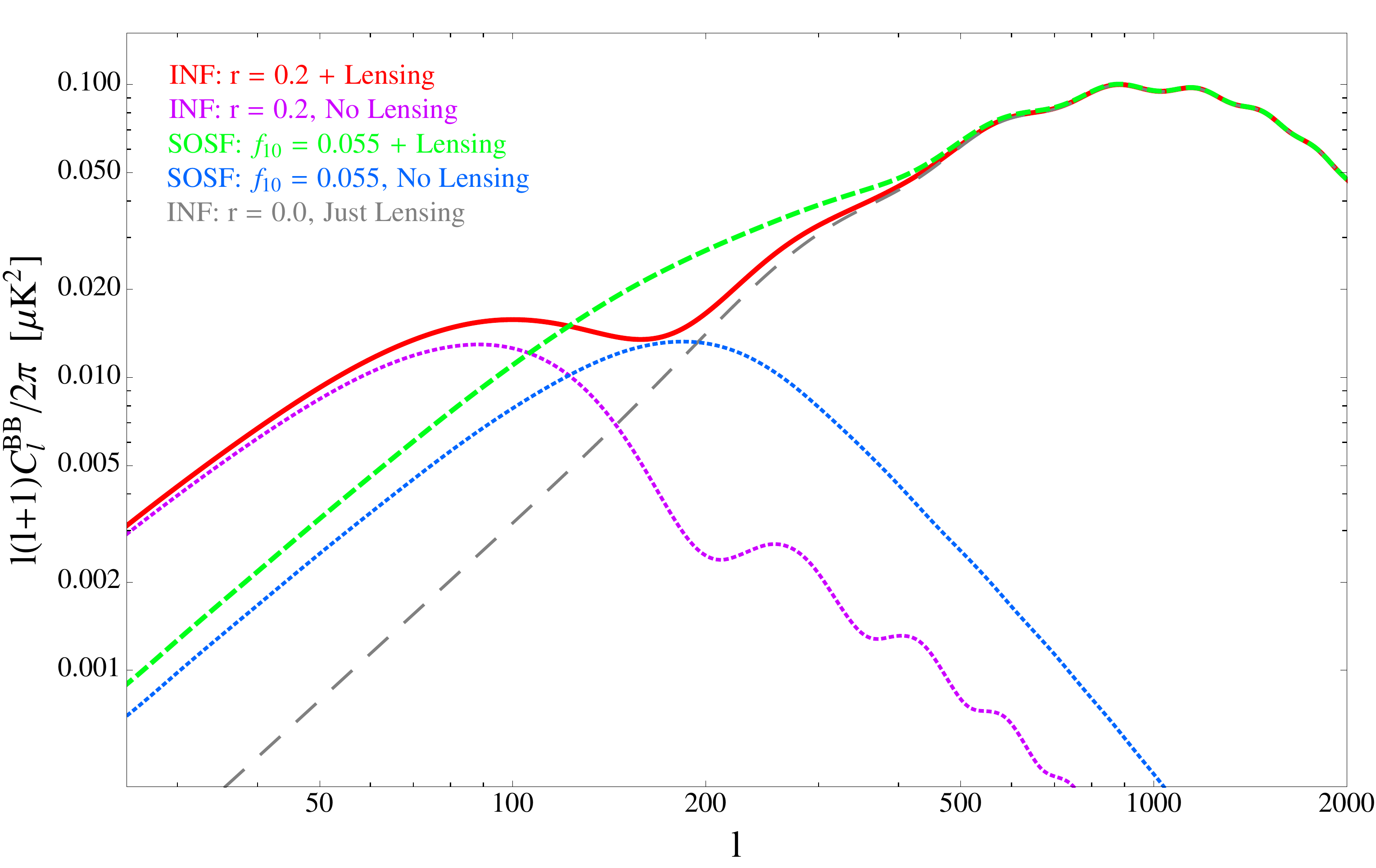}
\end{center}
\caption{SOSF B-mode angular power for $f_{10} = 0.055$ including lensing (green dashed line), which is the sum of the SOSF B-mode signal without lensing (blue dotted line) plus the lensing E-modes (gray dashed line). The analogous B-mode angular power from inflation for $r= 0.2$ with lensing is also shown (red continuous line); this is the sum of the inflationary signal without lensing (pink dotted line) and the lensing E-modes.}
\label{fig:2}
\end{figure}

Before showing the explicit results for the B-mode angular spectrum from the large-$N$, we emphasize a few aspects which highlight the difference of SOSF (and actually of any cosmic defect network) from inflation:\\

i) SOSF, like any network of cosmic defects, generate all type of perturbations, scalar, vector and tensor. Inflation, on the contrary, generates only scalar and tensor perturbations\footnote{As long as gravity is described by general relativity, even if vector perturbations are created during inflation, they subsequently decay due to the expansion of the Universe.}. 
In Figure~\ref{fig:1}, the B-mode angular power spectra from inflation and from SOSF are shown; the former only contains a tensor mode whereas the latter has both tensor and vector contributions. \\

ii) SOSF generate metric perturbations  actively throughout the evolution of the Universe. Inflationary perturbations, on the other hand, are produced during inflation, setting up initial conditions for the post-inflationary cosmic expansion. Because of this fundamental difference, SOSF are said to be an `active' seed mechanism, as opposed to inflation which lead to  `passive' perturbations.\\

iii) Metric fluctuations from SOSF (or other scaling sources) are typically white noise on super horizon scales (no correlations). This is a simple consequence of the linearized Einstein equations on those scales, which force metric perturbations to follow the energy-momentum power spectrum, which is white noise. For instance, tensors on super horizon scales are given by~\cite{Durrer:2001cg}
\begin{eqnarray}\label{eq:SuperHtentors}
h_{ij}(\bk,t) \simeq  \frac{1}{M_P^2\HH^2}T_{ij}^{\TT}(\bk,t)  \, , \qquad kt\ll 1\,,
\end{eqnarray}
where $T^{\TT}_{ij}$ denotes the tensor (spin-2) part of the energy momentum tensor of the large-$N$ scalar field and $M_P$ is the reduced Planck mass. Fluctuations from inflation are however scale invariant from the beginning right after inflation on all relevant scales.\\

iv) B-modes in the CMB polarization field can be created from lensing of E-modes by scalar perturbations, but mostly on small angular scales, i.e.~large $\ell$ (with a peak at $\ell \sim 1000$, see gray dashed line in Figure~\ref{fig:2}). Low multipole B-modes are mainly generated directly, from tensor and/or vector perturbations, but not from scalars~\cite{Kamionkowski:1996zd,Seljak:1996gy}. The low-$\ell$ B-modes detected by BICEP2, if confirmed, imply therefore that either tensor or vector perturbations, or a combination of both, have been detected. If the B-mode signal is to be interpreted within the inflationary framework, then tensor (and only tensor) perturbations are responsible, and thus gravitational waves have been (indirectly) detected. However, if the signal detected is attributed to SOSF, then both tensor and vector perturbations contribute. \\

v) Let us suppose that the BICEP2 signal could be entirely due to SOSF tensor perturbations, and the low-$\ell$ B-modes could be purely due to GW from SOSF. As mentioned in section~\ref{sec:intro}, any network of cosmic defects emits a scale invariant background of GW~\cite{Figueroa:2012kw,Krauss:1991qu,JonesSmith:2007ne,Fenu:2009qf, Giblin:2011yh}, so that one might (erroneously) expect that this would mimic an inflationary B-mode (with zero tensor tilt). However this is not the case, since the GW from SOSF (and defects in general) on the last scattering surface are scale invariant only for modes which are already inside the horizon. Hence, any tensor modes from SOSF which could be responsible for the low-$\ell$ BICEP2 B-signal, are not scale invariant, but closer to white noise a the time of decoupling, which yields $\ell(\ell+1)C_\ell^B \propto \ell^2$ over the relevant scales\footnote{The spectrum is not quite $\ell^2$ as it turns into scale-invariant at the horizon scale corresponding to $\ell\sim 150$.}, and therefore would not mimic an inflationary signal, see Fig.~\ref{fig:1}. The confusion of a GW backgrounds from inflation and SOSF can occur only  for direct detection GW experiments, but not in the B-polarization of the CMB. The difference between the inflationary and a defect  GW signal in the CMB has also been studied quantitativly in~\cite{Urrestilla:2008jv,Mukherjee:2010ve}.\\

Let us now move to our results. In Fig.~\ref{fig:1} we show the B-mode angular power spectra from inflation assuming $r = 0.2$ and zero spectral tensor tilt $n_t = 0$ (purple continuous), and from SOSF assuming a fractional contribution $f_{10} = 0.055$ (blue continuous). The first noticeable feature is that the inflationary curve is peaked at $\ell \simeq 90$, whereas the SOSF one is peaked at  roughly the double, $\ell \simeq 185$. Furthermore, the inflationary B-modes show oscillations at $\ell \gtrsim 200$, whereas those from SOSF do not. We could also note the difference in the very low-$\ell$  ($\ell\lesssim 15$) tail of the B-power spectra due to reionization. However, since BICEP2 has not measured these multipoles, we  ignore this difference here, and when plotting B-mode power spectra we  start from $\ell \gtrsim 10$.  We finally remark that  the inflationary B-spectrum comes only from tensors, while to the SOSF B-signal, both tensor and vector perturbations contribute, as indicated by the dashed lines in the figure (red and green dashed lines for the vectors and for the tensors respectively in Fig.~\ref{fig:1}). Vector perturbations create even a bigger signal than tensors for all the relevant scales outside the small interval $\ell \in [15,58]$. The low multipole data points of BICEP2 are at $\ell \sim 45, 74, 109$ and $144$ (central values), so if the BICEP2 signal was attributed to SOSF only, the amplitudes at multipoles $\ell \sim 74, 109$ and $144$ would be dominated by the vector contribution, whereas at $\ell \sim 45$ there would be a mix from tensors and vectors. Therefore, a first conclusion from this analysis is that if the BICEP2 signal was due (and only due) to SOSF, this would not imply a very strong detection of GW. It would  instead represent mainly a detection of vector perturbations (although in SOSF models generically both vector and tensor modes contribute). Anticipating our results, we will see below that, in reality, only a very small fraction of the BICEP2 signal can be due to SOSF, and therefore these conclusions do not hold.

\begin{figure}[t]
\begin{center}
\includegraphics[width=12cm]{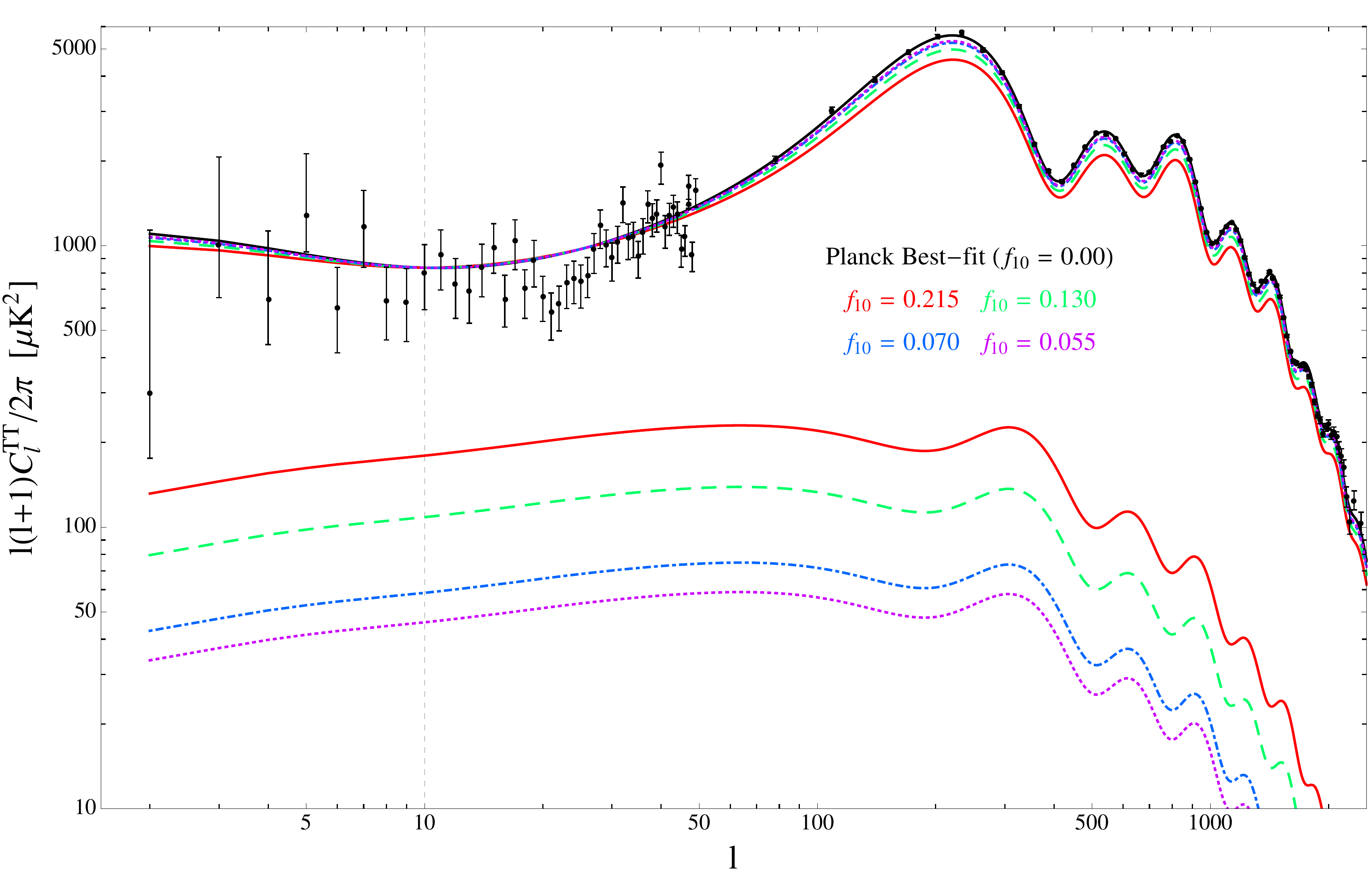}
\includegraphics[width=7.5cm]{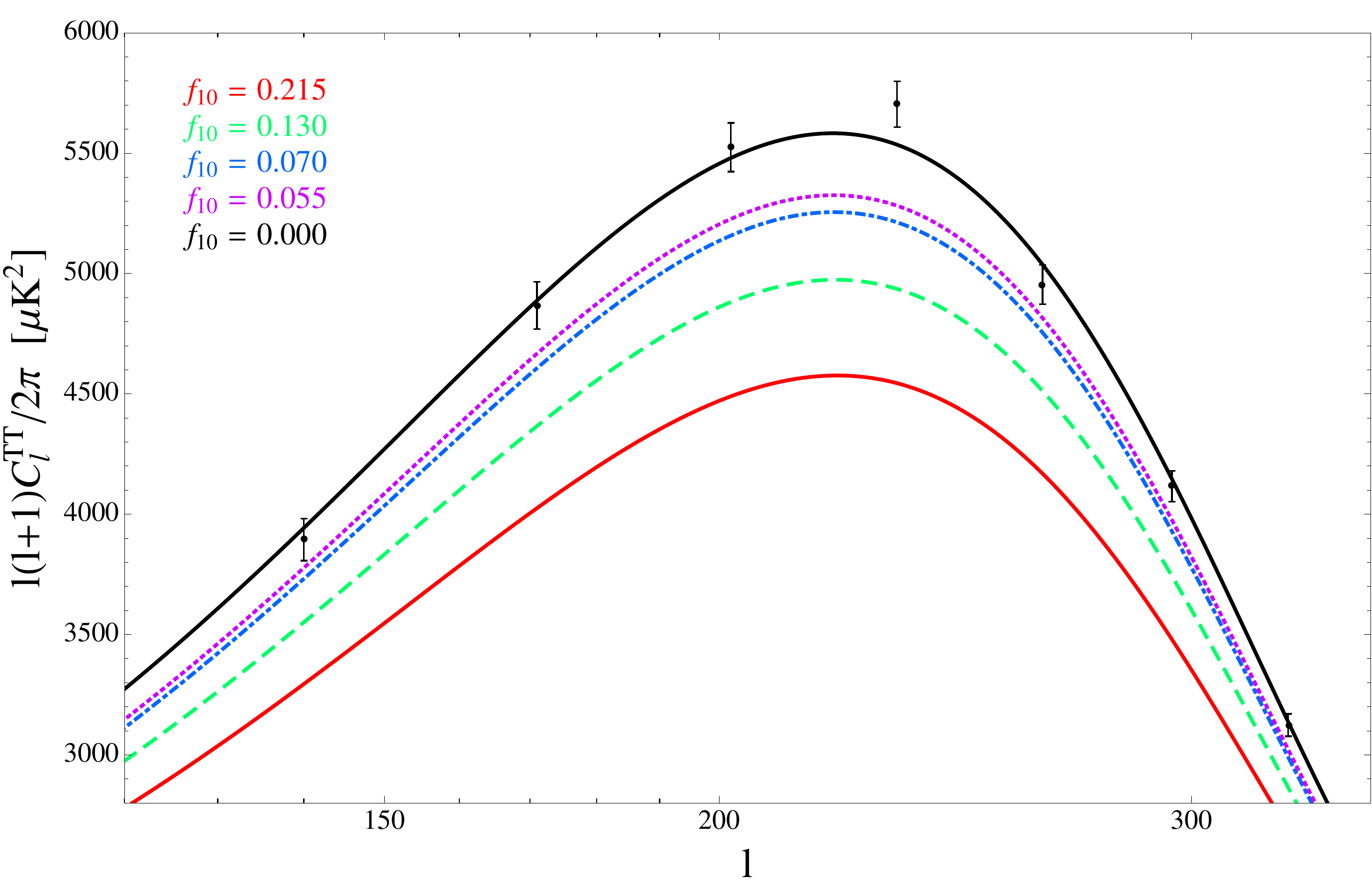}
\includegraphics[width=7.5cm]{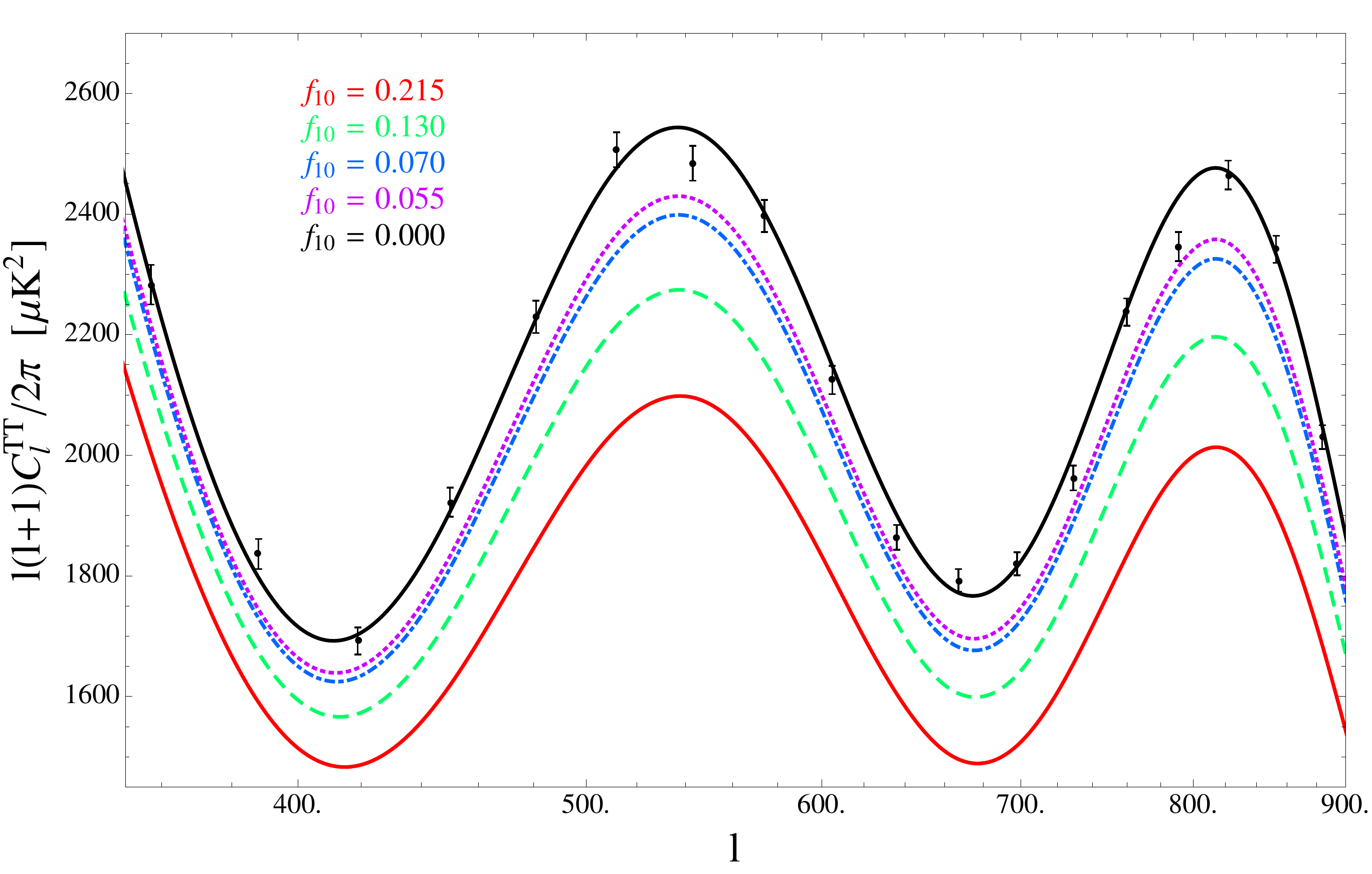}
\end{center}
\caption{\underline{Big panel}: Temperature angular power spectrum of scenarios combining inflation + SOSF, with different contributions from SOSF (top curves). Also shown are the signals from SOSF alone (lower curves). \underline{Small panels}: Zoom on the first acoustic peak (left), and 2nd and 3rd acoustic peaks (right) for the models shown in the big panel. The data points and errors  are from the {\it Planck} data. The different values of $f_{10}$ correspond to the attempt to fit the BICEP2 data with only the SOSF B-mode signal, see Fig.~\ref{fig:4}.}
\label{fig:3}
\end{figure}

In Fig.~\ref{fig:2} we decompose the total B-power spectra into primordial plus lensing contributions, both in the case of inflation and SOSF.  The shape of the SOSF + inflationary scalar B-mode spectrum shown in (green, dashed) is clearly very different from the pure inflationary tensor + scalar spectrum (solid, red). Especially, the `blue' SOSF spectrum $\sim \ell^2$ does not show the characteristic `plateau' at $70 \lesssim \ell \lesssim 180 $ which has been measured by BICEP2, and is well reproduced by the inflationary scenario. 

Let us now 
define the fractional contribution of the SOSF (and of cosmic defects in general) to the temperature anisotropies at multipole $\ell = 10$ as 
$$f_{10} \equiv {C_{10}^\TT\big|_{\rm SOSF}\over C_{10}^\TT\big|_{\rm obs}}\,,$$
where $C_\ell^\TT\big|_{\rm x}$ are the temperature angular power spectra from x = SOSF and x= obs the spectrum observed by {\em Planck} (modeled by an inflationary signal). The analysis of the {\it Planck} collaboration indicates that $f_{10} \lesssim 0.015, 0.03, 0.045$ for Nambu-Goto, Abelian-Higgs and semi-local strings, respectively, and $f_{10} \lesssim 0.055$ for O(4)-global textures. Unfortunately, there are no upper bounds set for $f_{10}$ for the large-$N$ limit of SOSF's, neither from {\it Planck} nor from the successive series of WMAP 1-, 3-, 5-, 7 and 9-year analysis. However, the O(4)-global textures studied by {\it Planck} already capture well the large-N limit of SOSF, since with temperature anisotropy spectra normalized at a given multipole (say $\ell = 10$), their B- power spectra only differ by a few $\%$. Therefore, it should suffice to take the upper bound $f_{10} \lesssim 0.055$ from O(4)-global textures as an approximate upper bound $f_{10}$ for SOSF, and significantly bigger values of $f_{10}$ are probably ruled out by the {\it Planck} temperature data. One could expect a slightly different upper bound for $f_{10}$ for the large-N SOSF case due to the different spectra, and being generous we will allow for bigger values than $0.055$. But as we shall see later, the BICEP2 data actually decrease the upper bound of $f_{10}$ for the large-N SOSF well below the $0.055$ limit for O(4)-global textures. For the time being we  assume $f_{10} = 0.055$ as a reference value.

\begin{figure}[t]
\begin{center}
\includegraphics[width=12cm]{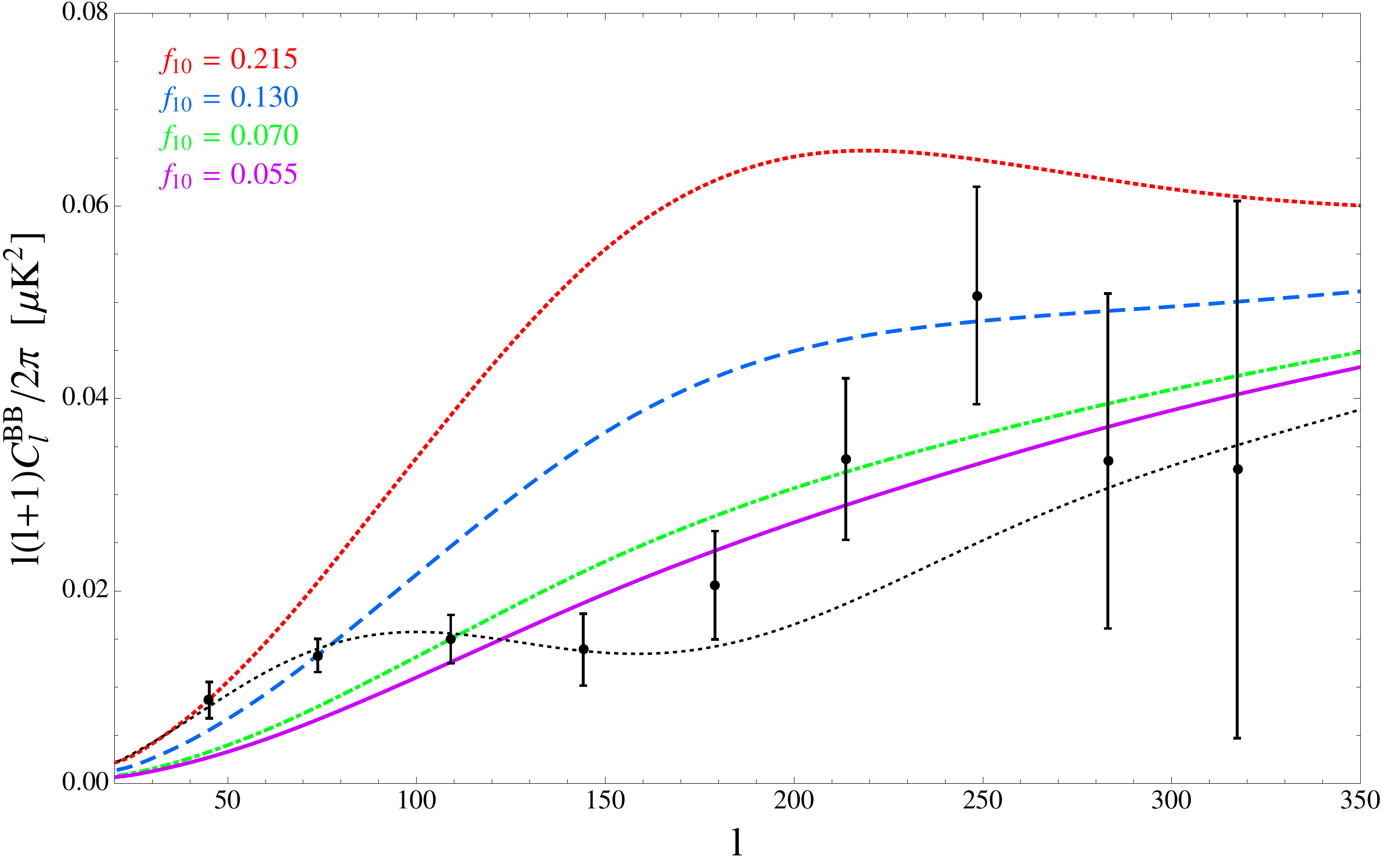}
\end{center}
\caption{BICEP2 data points with errors and the B-mode angular power spectra from SOSF's, choosing the  values $f_{10}$ such that the spectra pass respectively through the first ($f_{10} = 0.215$), second ($f_{10} = 0.130$) and third ($f_{10} = 0.070$) data point. Also shown is the spectrum for $f_{10} = 0.055$. Clearly, fitting the BICEP2 data purely with SOSF is impossible, since the spectra simply have the wrong shape.}
\label{fig:4}
\end{figure}

\begin{figure}[t]
\begin{center}
\includegraphics[width=7.5cm]{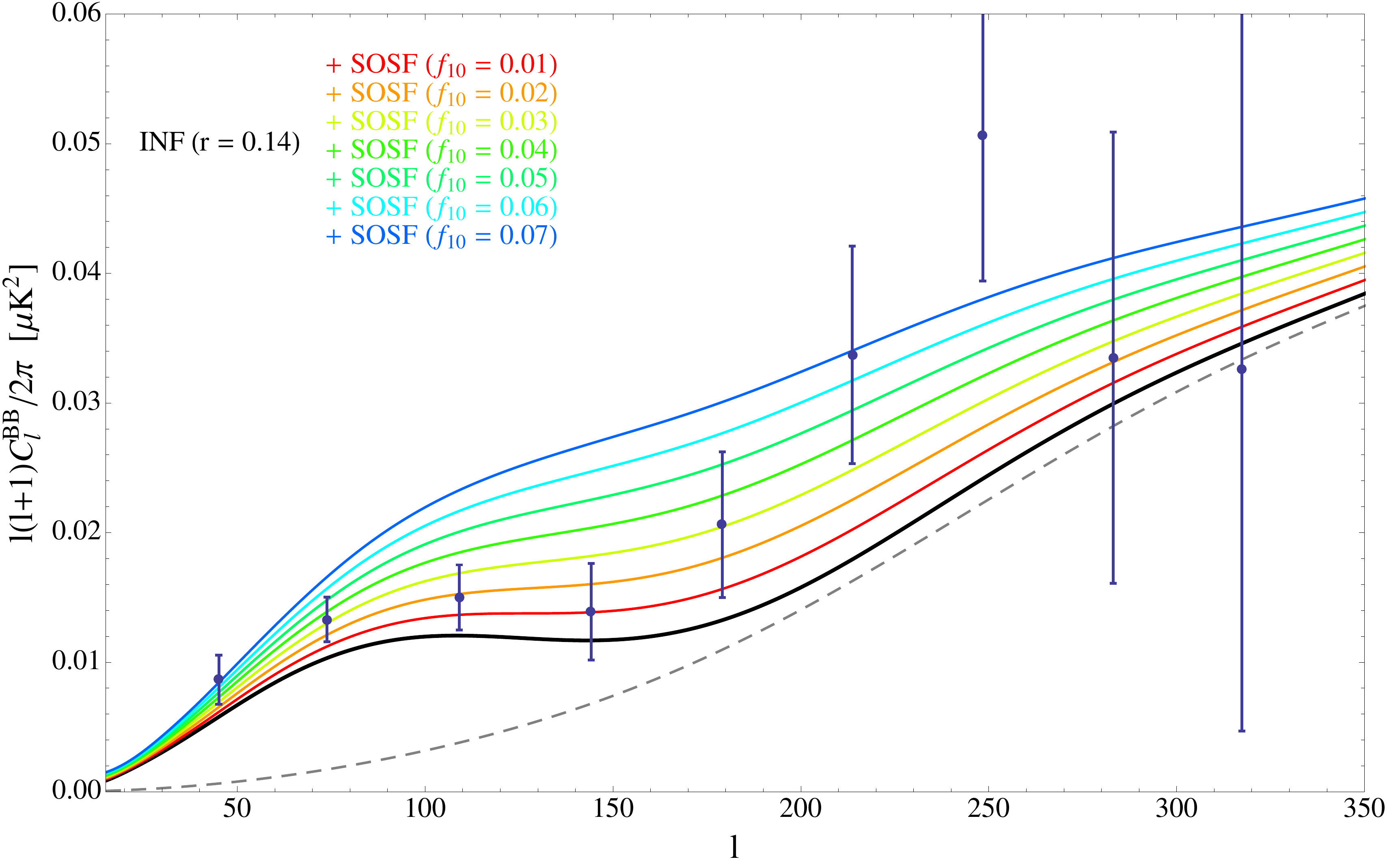}
\includegraphics[width=7.5cm]{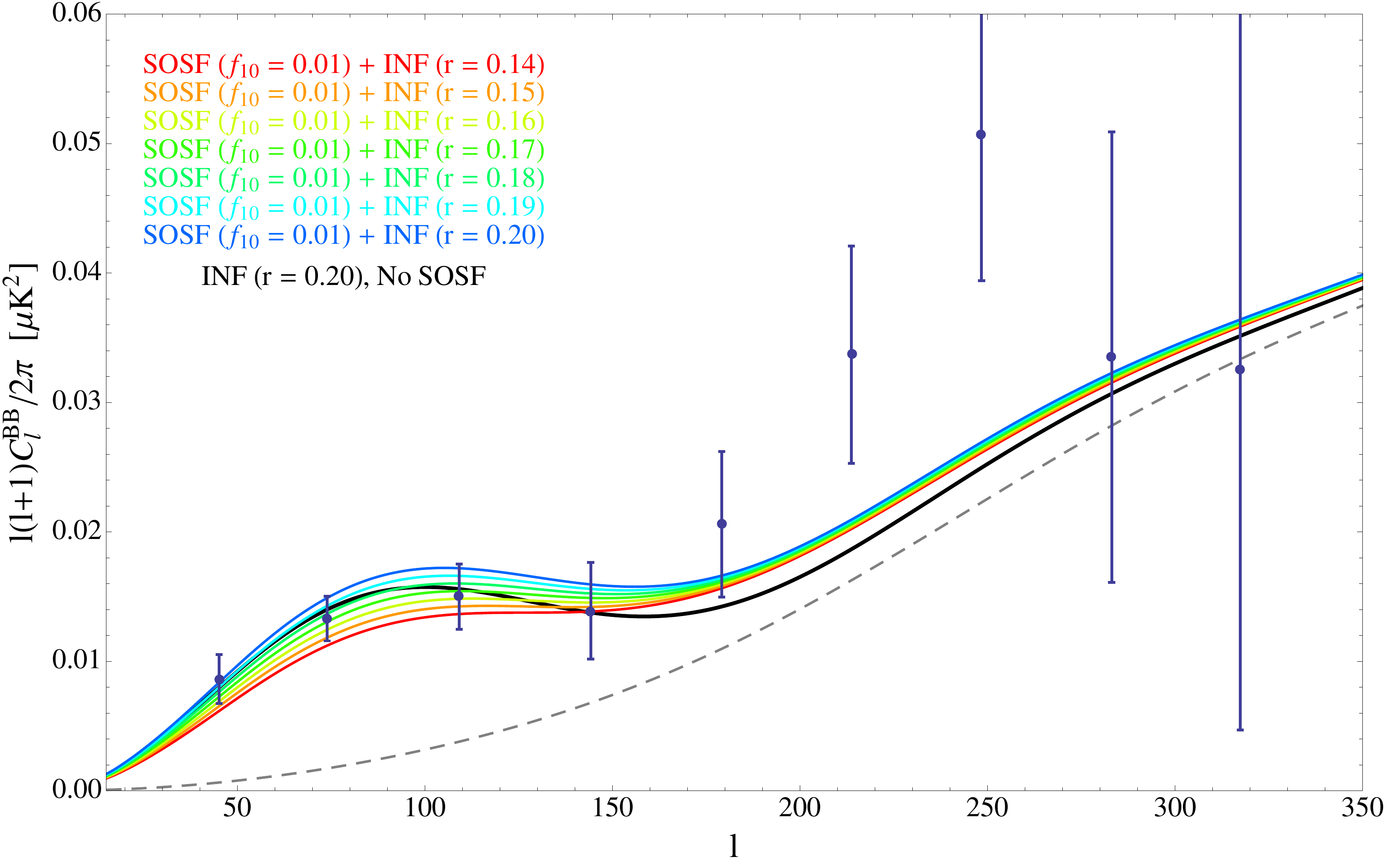}
\includegraphics[width=7.5cm]{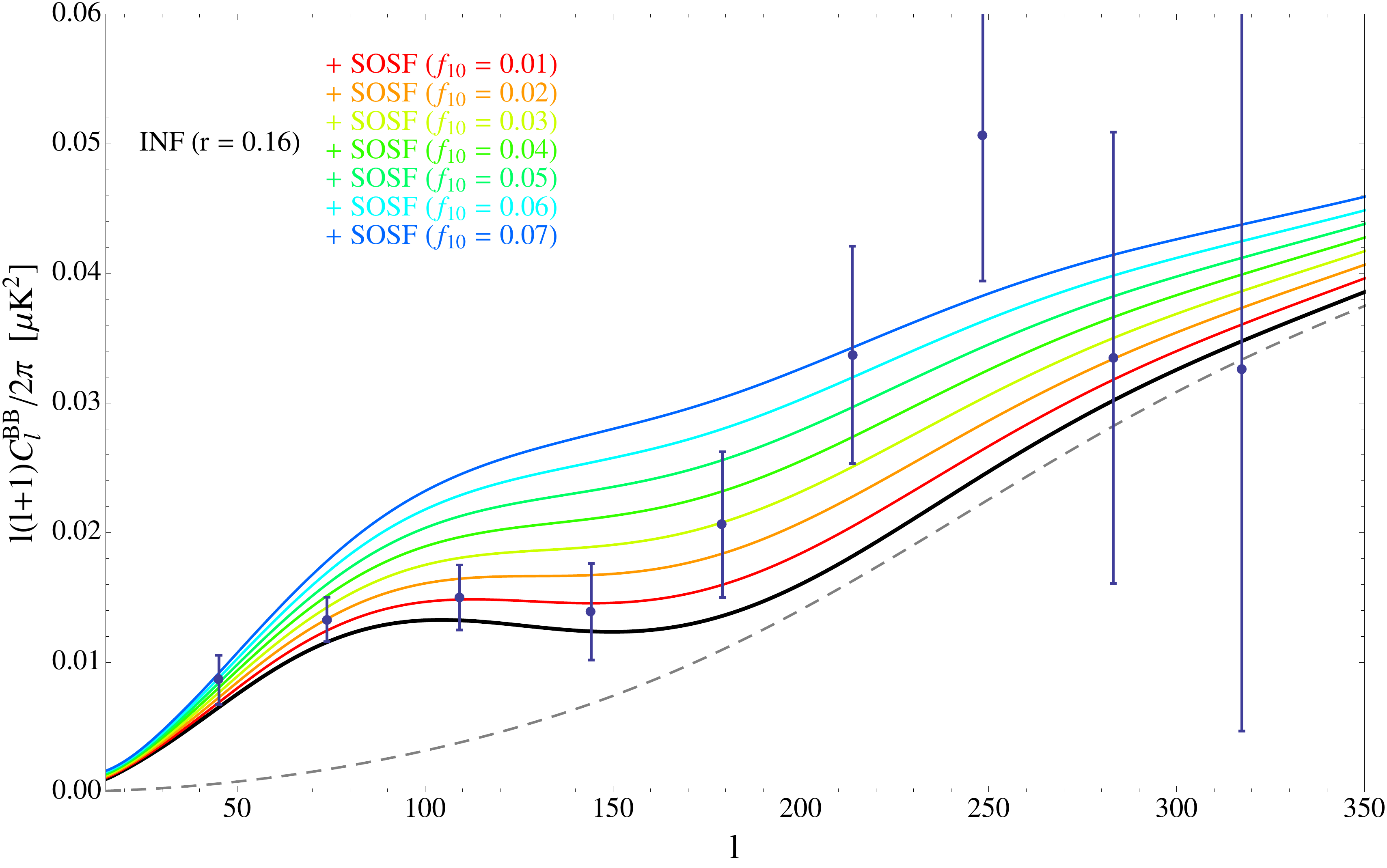}
\includegraphics[width=7.5cm]{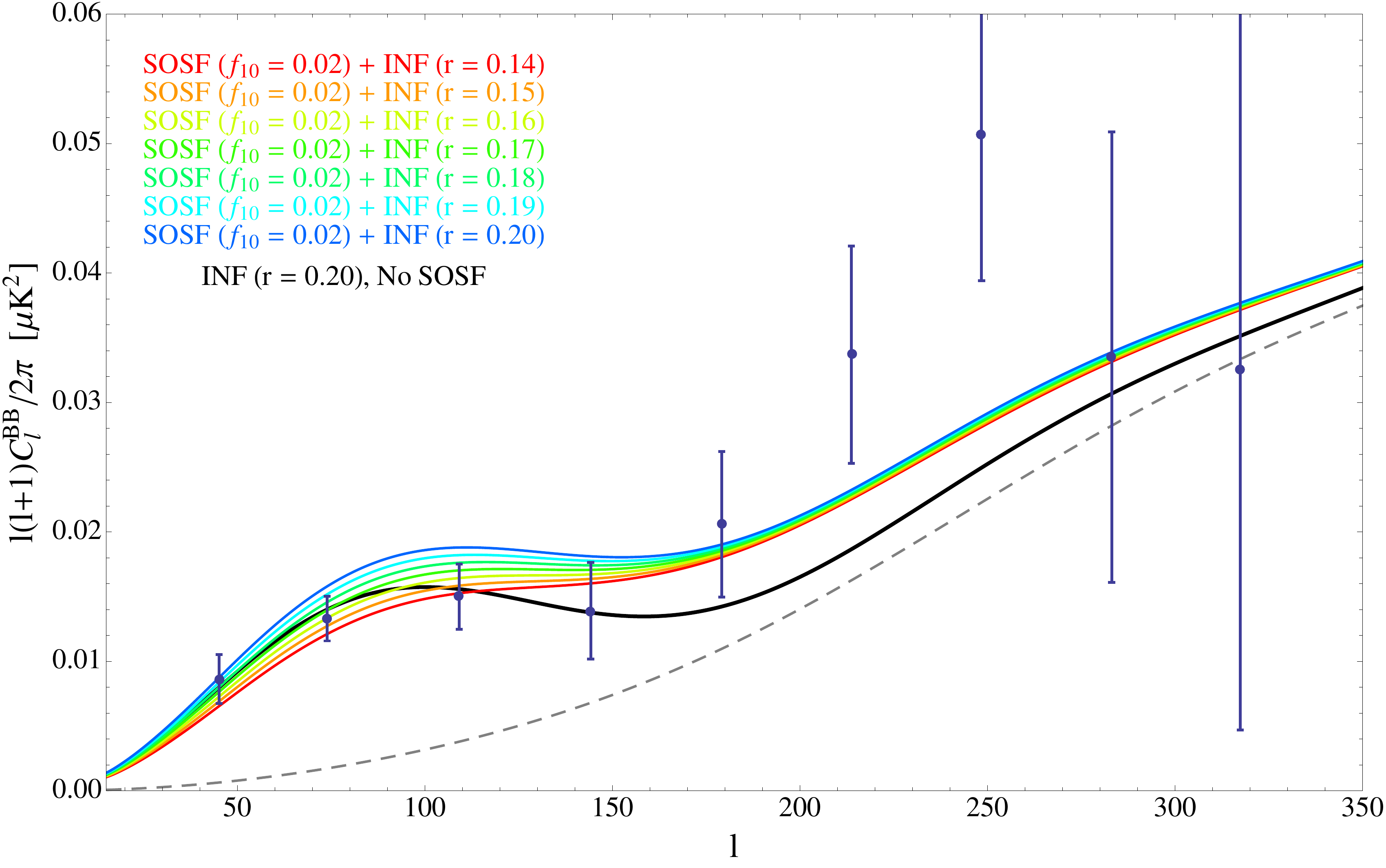}
\includegraphics[width=7.5cm]{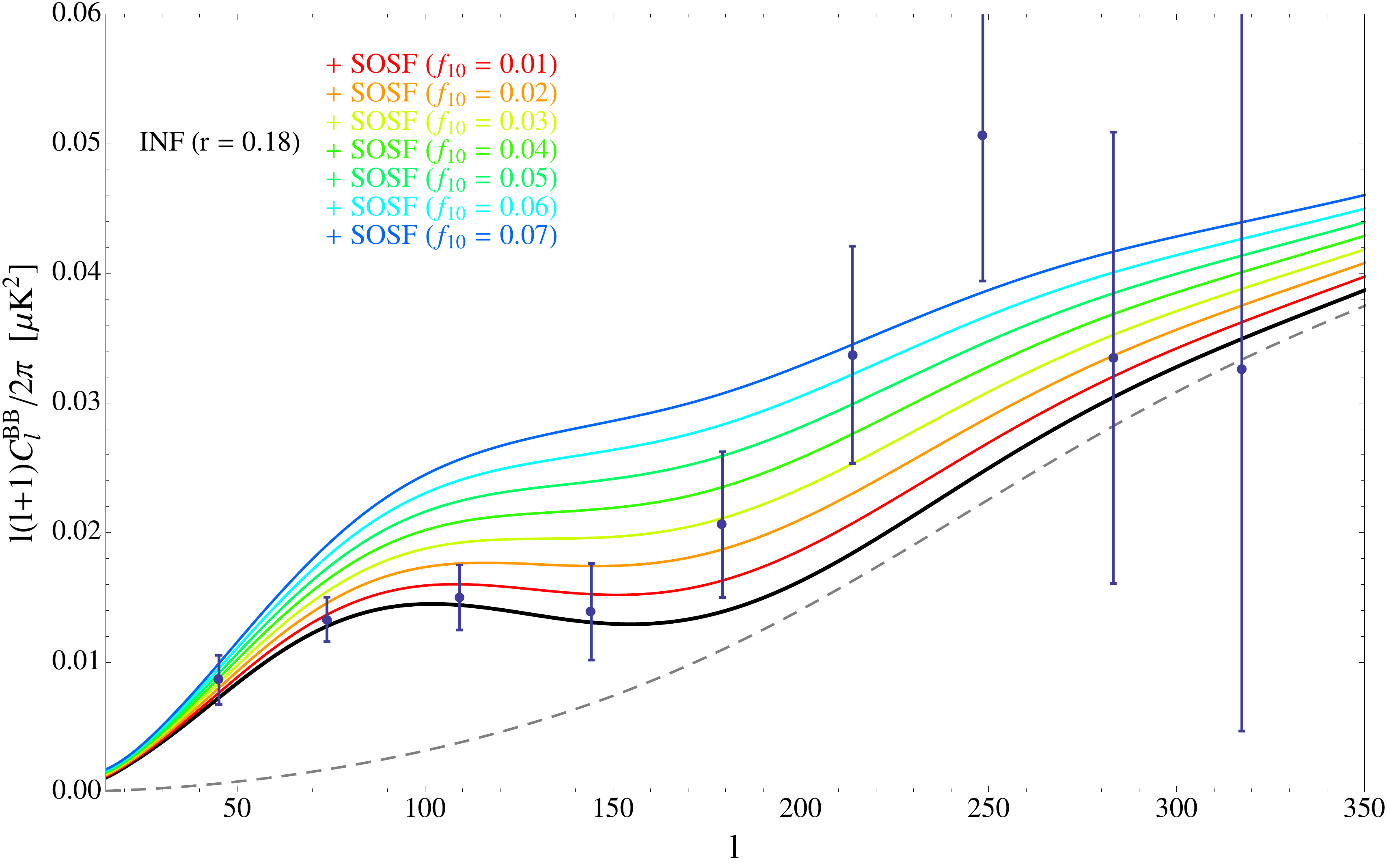}
\includegraphics[width=7.5cm]{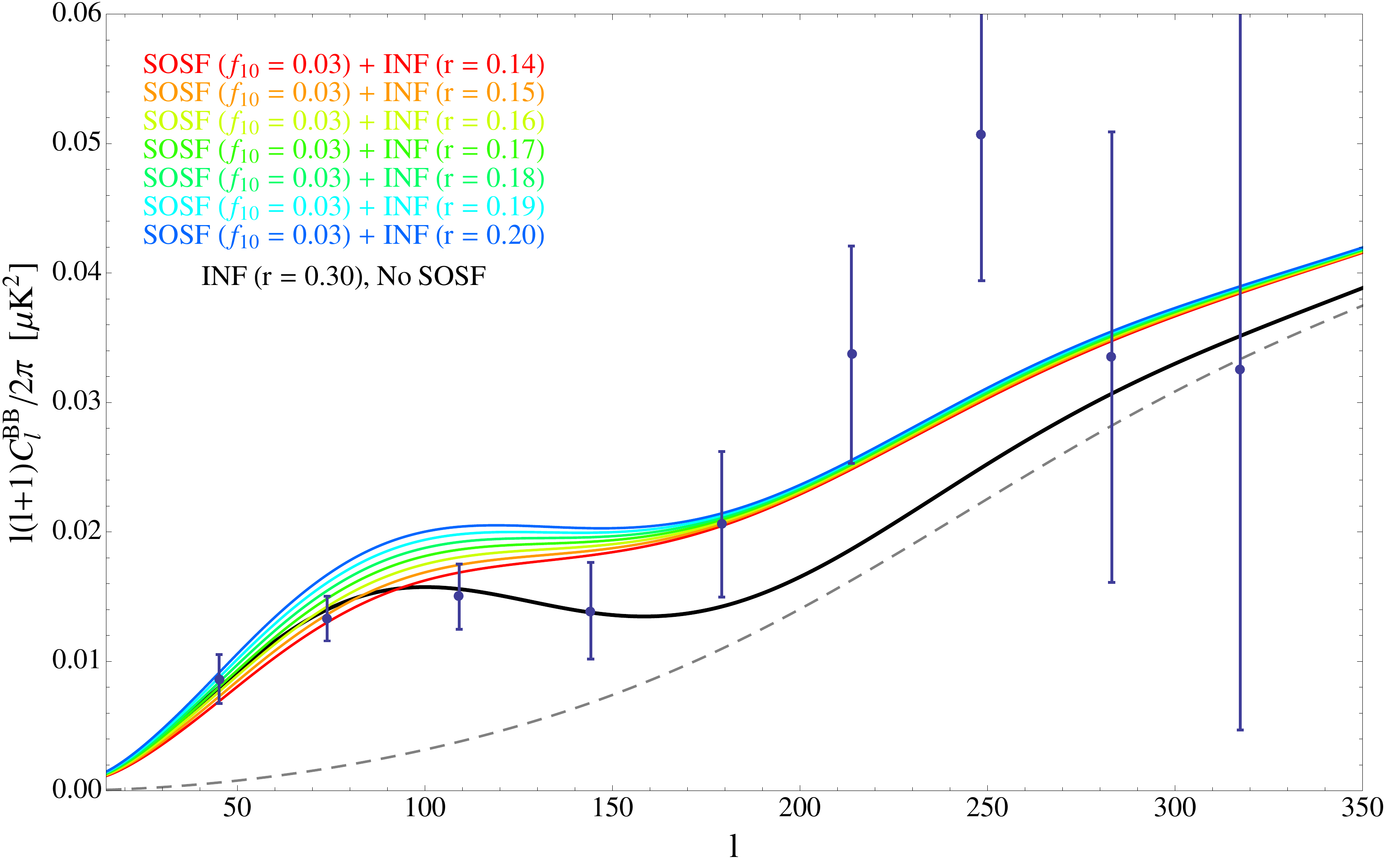}
\end{center}
\caption{B-mode angular power spectrum, Left: fixing inflationary $r$ and varying $f_{10}$, Right: fixing $f_{10}$ and varying inflationary $r$. The chosen values for $r$ and $f_{10}$ are given in the figures. The BICEP2 data are also indicated.}
\label{fig:5}
\end{figure}

In Fig.~\ref{fig:3} we show the temperature power spectrum from the {\it Planck} data. We plot the {\it Planck} best-fit in black and the {\it Planck} best-fit for different fractional contribution from the SOSF  in different colors as indicated in the figure. Zooming into the first or the second and third acoustic peaks (lower panels), we see that a contribution from SOSF, when normalized such that the low-$\ell$ Sachs-Wolfe plateau remains unchanged, reduces the amplitude of the acoustic peaks.
Note that this is not a MCMC best fit including defects but just an addition of the defect component to the inflationary scalar component with adjustment of solely the amplitude, while all other cosmological parameters are fixed to the {\it Planck} best fit values. Nonetheless, these plots indicate very clearly that the {\it Planck} temperature data does not favor a contribution from SOSF.

In Fig.~\ref{fig:4} we attempt different fits to the B-mode BICEP2 data with only SOSF + lensing. We fix $f_{10}$ to $0.215,~0.130,~0.070$ and $0.055$ such that the curves pass successively though the 1st, 2nd and 3rd BICEP2 data points. We also show the B-mode angular spectrum for the case $f_{10} = 0.055$ (which does not pass though any of the BICEP2 points), since as we mentioned before this represents the maximum fraction allowed by {\it Planck} (at least for an O(4)-global texture). Clearly, no choice for $f_{10}$ yields a good fit. The data simply have a different shape and are much better fitted by an inflationary spectrum (indicated with a black dotted line). Besides, even if we were able to somehow fit the B-mode signal alone with SOSF, the required values for $f_{10}$ are so big that clearly they would be in tension with the temperature angular spectrum measured by {\it Planck}, as indicated by Fig.~\ref{fig:3}.

In Fig.~\ref{fig:5} we finally vary both $r$ and $f_{10}$ at the same time. On the left panels we vary $f_{10}$ for three given values of $r$, while in the right panels we vary $r$ for three given values of $f_{10}$. The BICEP2 data show that a combination of an inflationary signal with $r = 0.16$ and $f_{10} = 0.01-0.02$ is a good fit to the data, alleviating the tension with {\it Planck} and the deviation of three of the data points from BICEP2 at higher $\ell$'s. The presence of SOSF indeed improves the fit to the B-mode BICEP2 data, but clearly points towards a marginal contribution of the SOSF. Note, however, that this conclusions are not the result of a rigorous MCMC analysis, but come just from looking at the variation of $r$ and $f_{10}$ in Fig.~\ref{fig:5}. At the present level of B-mode data, this is still reasonable. From the left panels in Fig.~\ref{fig:5} we clearly see that, independently of the value of $r$, the low-$\ell$ BICEP2 data is not compatible with $f_{10} > 0.03$. From the right panels in Fig.~\ref{fig:5} we clearly see that, when considering an admixture scenario of inflation + SOSF, the BICEP2 data prefer a value of the inflationary tensor-to-scalar smaller than 0.2, in the range $r \sim 0.15-0.17$.

\section{Discussion and Conclusions}
\label{sec:conclusions}

We have investigated the B-polarization of the CMB from a SOSF model and shown that it cannot reproduce the BICEP2 data.  This is due to the fact that the polarization signal comes directly from the last scattering surface at which the BICEP scales corresponding mainly to $50<\ell<150$, are still super-horizon. On the other hand, the super-horizon spectrum of the SOSF tensor modes is uncorrelated white noise and not scale invariant. The same causality argument implies that defects and also SOSF have no first peak in the TE spectrum as has been pointed out in Ref.~\cite{Spergel:1997vq}. 

In the same spirit, the low-$\ell$ B-polarization spectrum from defects and SOSF is not scale invariant but much bluer and can be used to set strong limits on a possible contribution from defects, here modeled as SOSF. We therefore conclude that SOSF cannot explain on their own the BICEP2 data.

Thanks to the BICEP2 data, we have also been able to limit the possible contribution from SOSF to the CMB signal to a marginal fraction of about $f_{10} \lesssim 0.02$. This shows the strength of B-modes for the discrimination of a contribution from scaling seeds which was already pointed out in~\cite{GarciaBellido:2010if}, see also~\cite{Lizarraga:2014eaa}. Future, more precise data will help us to constrain $f_{10}$ much better. For the time being, considering a mixed scenario of inflation + SOSF helps to improve the fit to the current BICEP2 data, in particular, it provides a better fit to the anomalous large amplitude of the high-$\ell$ multipoles. Nevertheless, adding SOSF also reduces the inflationary tensor-to-scalar ratio to $r \approx 0.16$, i.e.~towards a value that is better compatible with the upper bound from  {\it Planck} (in the absence of running of the scalar index).

Let us also note that this is actually the first constraint for the  contribution from the large-$N$ limit of SOSF to the CMB. Since, not surprisingly, the result is very similar to the one for textures ($N=4$), this allows us to generically constrain global $O(N)$ models to contribute at most a few percent to the BICEP2 data and to the CMB anisotropies and polarization in general. The same is expected to hold for other scaling scalar field models with similar characteristics. The main point here is that the energy momentum  tensor of the source is uncorrelated on super horizon scales, and that the only scale of the problem is the Hubble horizon scale.

\section*{Acknowledgements}

We are grateful to Mark Hindmarsh and Jon Urrestilla  for useful comments. We thank Elisa Fenu and Juan Garc\'ia-Bellido for their collaboration in a previous related project. We acknowledge financial support from the Swiss National Science Foundation.

\bibliography{CLNrefs}

\bibliographystyle{h-physrev4}

\end{document}